\documentclass[11pt]{article}

\usepackage[preprint]{acl}

\usepackage{times}
\usepackage{latexsym}

\usepackage[T1]{fontenc}

\usepackage[utf8]{inputenc}

\usepackage{microtype}

\usepackage{inconsolata}

\usepackage{graphicx}
\usepackage{booktabs}
\usepackage{multirow}
\usepackage{amsmath}
\usepackage{xcolor}
\usepackage{float}
\usepackage{enumitem}
\title{What Makes a Good Audio Prompt for Zero-Shot TTS?\\Learning Prompt Selection from Synthesis Outcomes}

\author{Arnon Turetzky \and Yossi Adi \\
  The Hebrew University of Jerusalem}

\begin{document}
\maketitle

\begin{abstract}
Zero-shot text-to-speech (TTS) models clone a target voice from a short reference clip, commonly referred to as the \emph{prompt}.
In practice, the prompt has to be cut from a longer recording of the speaker, so which part to use must be chosen.
To understand what makes a good prompt, we analyze hundreds of thousands of outputs synthesized by four TTS models from candidate prompts of YouTube recordings.
The choice of prompt substantially changes intelligibility, naturalness and speaker similarity.
Standard predicted-quality scores capture mainly the naturalness part but miss most of the intelligibility gain.
We find that the models largely agree on which prompt yields the highest measured speaker similarity, but much less on which prompt yields intelligible speech.
We propose a small network that learns from the outputs a TTS model produced to rank a recording's candidate prompts from their audio alone.
Without synthesizing anything at selection time, it picks prompts that lower the word error rate on held-out recordings by a third (Qwen3-TTS) to two fifths (F5-TTS) relative to a randomly chosen prompt, more than predicted quality achieves and nearly as much as synthesizing a dozen test sentences from every candidate first.
\end{abstract}

\section{Introduction}
\label{sec:intro}

\begin{figure}[t]
\centering
\includegraphics[width=\columnwidth]{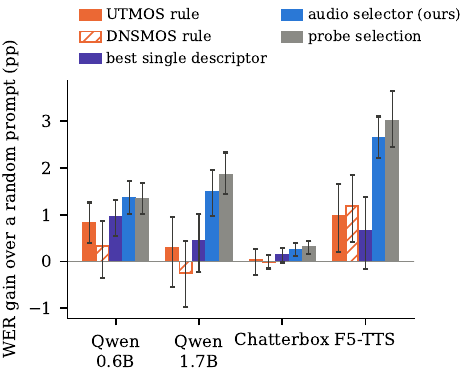}
\caption{Our audio selector, a small network trained on a TTS model's synthesis outcomes, picks prompts that lower WER more than choosing by predicted quality, without synthesizing anything. WER gain over a random prompt (pp, higher is better) on 400 held-out test recordings. Probe selection uses 12 syntheses per candidate, and Qwen3-TTS 1.7B uses the 0.6B selector and descriptor. Error bars: 95\% intervals. Details in \S\ref{sec:models} and Appendix~\ref{app:models}.}
\label{fig:models}
\end{figure}

Zero-shot TTS models can generate speech in a new voice from just a few seconds of that speaker's audio, the \emph{prompt} \citep{wang2023valle,anastassiou2024seedtts,roth2024hebtts,chen2024f5tts,du2024cosyvoice2,wang2025maskgct,meng2025melle,turetzky2025salad}.
In benchmarks each test sentence comes with its prompt, but in practice the prompt is not given: a user has a longer recording of the speaker, such as a video, a podcast or a voice message, and has to decide which few seconds to use.
We ask whether this choice matters, and how to make it well.

Prior work chooses prompts mainly to match a target style or emotion, or screens them by annotated or predicted quality (\S\ref{sec:related}).
Among one speaker's candidates, the common practice is to take the one with the best predicted audio quality, as scored by non-intrusive predictors such as DNSMOS and UTMOS \citep{saeki2022utmos}, standard tools for filtering TTS training data \citep{he2024emilia} and evaluating synthesized speech \citep{chen2024f5tts}.
Yet a prompt that sounds clean need not be the one from which a model speaks clearly, and it remains unclear whether quality scores rank a speaker's candidate prompts by the speech a TTS model produces from them.

To find out, we synthesize speech from every candidate prompt of over 1,500 YouTube recordings with four TTS models, and measure how intelligible, natural and similar to the speaker each output is.
The choice turns out to matter on every metric, and not only by chance: part of the difference is a stable property of the prompt.
Choosing by predicted quality, however, captures mainly the naturalness part and misses most of the gain in intelligibility.

Synthesizing from every candidate tells us which prompt is good, but it is too costly to repeat for every new speaker.
We therefore learn to predict it: a small network that hears only the candidates' audio is trained to rank them by the outputs a TTS model produced from them.
On held-out recordings, its choice lowers word error rate by a third to two fifths relative to a randomly chosen prompt, without synthesizing anything (Table~\ref{tab:final}, Figure~\ref{fig:models}).
This is more than the quality-based choice gains, and close to what synthesizing a dozen test sentences from every candidate would find.

\paragraph{Our contributions:}
\begin{itemize}[leftmargin=*,nosep]
\item We frame prompt selection as choosing, among a speaker's candidate prompts, the one that leads to the best speech, and measure it at scale for four TTS models (\S\ref{sec:setup}).
\item We find that models largely agree on which prompt yields the highest measured speaker similarity but much less on which yields intelligible speech, and that choosing by predicted quality improves mainly naturalness (\S\ref{sec:sensitivity}, \S\ref{sec:models}, \S\ref{sec:noise}).
\item We train a small neural network that ranks a speaker's candidate prompts from their audio alone, learning from the quality of speech a TTS model synthesized from other prompts, and that picks better prompts without synthesizing anything (\S\ref{sec:learned}, \S\ref{sec:analysis}).
\end{itemize}
We will release our code upon acceptance.
\section{Related Work}
\label{sec:related}

Zero-shot TTS is evaluated with one given prompt per test sentence \citep{anastassiou2024seedtts}; recent benchmarks vary the prompt's acoustic condition (clean, noisy, in the wild, expressive) to test the robustness of the models \citep{raon2026opentts}.
Prompt choice has received attention mainly for style and emotion: prompts are retrieved whose text or style matches the target sentence \citep{gong2021multiref,luo2025autostyle}, or screened for emotional expressiveness, audio quality and fit with the text, tested with the TTS model and matched to the input text \citep{wang2024expressive}; M$^3$PDB selects prompts from a database by annotated labels \citep{zhu2025m3pdb}.
For in-the-wild data, \citet{giraldo2026bsc} enhance noisy prompts and find that prompt quality and length affect the output, and \citet{sun2026dps} filter prompts with scores from audio and text language models.
Another way to avoid bad outputs is to synthesize several and keep the best by a learned score \citep{betker2023tortoise}, which multiplies the cost of every sentence.
We ask a different question: among prompts cut from the same recording, which one makes a fixed TTS model produce good speech for sentences not known in advance, measured by synthesis outcomes; and we learn a selector from those outcomes, once per TTS model, then make a single prompt choice per speaker.

Non-intrusive predictors estimate the naturalness or quality of speech from the audio alone \citep{saeki2022utmos,reddy2022dnsmos,tjandra2025audiobox}; they filter TTS training data \citep{he2024emilia} and evaluate TTS systems \citep{chen2024f5tts}.
We show where they fall short as a rule for choosing prompts, and how they flatter such a rule when they also judge it.

For codec language models, voice cloning has been described as in-context learning \citep{wang2023valle}.
In text, in-context learning is sensitive to the choice and order of the examples \citep{zhao2021calibrate,lu2022fantastically}, and examples are selected or learned from downstream outcomes \citep{liu2022makes,rubin2022learning}; for speech recognition with multimodal language models, in-context examples are retrieved by text similarity \citep{ticl2025}.
Our selector likewise learns from downstream outcomes which prompt to use.
\section{Setup}
\label{sec:setup}

\begin{table*}[t]
\centering\small
\begin{tabular}{lrrrrrr}
\toprule
selector & $\Delta$WER & $\Delta$fail. & $\Delta$UTMOS & $\Delta$DNSMOS & $\Delta$Audiobox & $\Delta$sim. \\
\midrule
UTMOS rule & 0.83 & 0.81 & 0.095 & 0.024 & 0.035 & -0.006 \\
DNSMOS rule & 0.33 & 0.54 & 0.038$^\ddagger$ & 0.032 & 0.019 & 0.027$^\dagger$ \\
best single descriptor (transcript confidence) & 0.96 & 0.55 & -0.002$^\ddagger$ & 0.006$^\ddagger$ & -0.010$^\ddagger$ & 0.017$^\dagger$ \\
\midrule
descriptor ranker & 1.00 & 0.41 & 0.069 & 0.028 & 0.033 & -0.007 \\
audio selector (ours) & \textbf{1.36$^\dagger$} & \textbf{0.87} & \textbf{0.050$^\ddagger$} & \textbf{0.014} & \textbf{0.022} & \textbf{-0.002} \\
\quad + UTMOS head ($w=0.25$, \S\ref{sec:learned}) & 1.33$^\dagger$ & 0.67 & 0.060$^\ddagger$ & 0.019 & 0.038 & -0.004 \\
\midrule
probe selection (12 syntheses per candidate) & 1.35$^\dagger$ & 0.88 & 0.206$^\dagger$ & 0.098$^\dagger$ & 0.282$^\dagger$ & 0.091$^\dagger$ \\
\bottomrule
\end{tabular}
\caption{Test results for Qwen3-TTS 0.6B (400 held-out recordings, all 19 candidates of a recording in one selection). $\Delta$: improvement over a randomly chosen candidate of the same recording ($\uparrow$ better; for WER and failures, the reduction in pp); Audiobox is the held-out judge. $^\dagger$/$^\ddagger$: better/worse than the UTMOS rule (interval of the difference above/below zero).}
\label{tab:final}
\end{table*}

\paragraph{Task.}
Each speaker comes with one \emph{recording}, from which a set of \emph{candidate prompts} is cut.
A selector picks one candidate before any synthesis. It may use the candidates' audio and automatic \emph{prompt transcripts}; our audio selector uses the audio alone, and the chosen prompt's transcript is then given to the TTS models that read one. As in practice, the prompt is chosen once and then used for any sentence, so the selector does not see the sentences that will be synthesized.
We judge a prompt by the speech it leads to: its \emph{outcome} on a metric is the average over the outputs synthesized from it.
We judge a selector by its \emph{gain}: how much better its pick's outcome is than that of a randomly chosen candidate (the mean over the recording's candidates).

\paragraph{Recordings and prompts.}
The recordings are English YouTube videos from YODAS \citep{li2023yodas} (CC-BY).
Since many videos contain several voices (interviews, hosts and guests), we keep only the utterances of a video's main speaker, identified by speaker embeddings, and only videos that this speaker dominates; we also require English speech, 3--30 minutes of audio and, since some videos have low-bandwidth, telephone-like audio, full-band audio (details in Appendix~\ref{app:repro}).
In each recording, a random 30\,s stretch that is mostly the main speaker, so that prompts rarely contain another voice, is cut, at equal total audio, into 10 prompts of 3\,s, 6 of 5\,s and 3 of 10\,s, at random positions with edges snapped to low energy; two prompts of the same length share at most half of their audio.
All 19 candidates of a recording compete in one selection, since a user could choose any of them.
Five \emph{target sentences} (4--10\,s) and three anchor utterances, against which speaker similarity is measured (rather than against the prompt itself), come from the same recording, outside the 30\,s stretch.
All transcripts are produced automatically by Whisper large-v3 \citep{radford2023whisper}, as a user without transcripts would.
We split the recordings into training, development and test sets with no overlap (Table~\ref{tab:data}).

\paragraph{TTS models and outcomes.}
We use four open zero-shot TTS models from three families with their default inference settings:
Qwen3-TTS \citep{qwen3tts} (0.6B and 1.7B parameters), an autoregressive codec language model that continues the prompt and reads its transcript;
F5-TTS \citep{chen2024f5tts} (0.3B), a non-autoregressive flow-matching model that also reads the prompt transcript;
and Chatterbox \citep{chatterbox} (0.5B), which conditions on the prompt audio only.
Every candidate prompt is used for the recording's five target sentences with three sampling seeds each, and every output is scored for
\emph{intelligibility}, by the word error rate (WER) of Whisper large-v3 against the target sentence, and by the \emph{failure rate}, the share of outputs with a WER of at least 50\% or a runaway generation;
\emph{naturalness}, by UTMOS \citep{saeki2022utmos}, DNSMOS P.835 \citep{reddy2022dnsmos} and Audiobox Aesthetics production quality \citep{tjandra2025audiobox}, which no selector uses and which serves as a held-out judge;
and \emph{speaker similarity}, by the cosine between ECAPA-TDNN speaker embeddings \citep{desplanques2020ecapa} of the output and of the anchors.
A prompt's outcome on a metric is its mean over its 15 outputs.
WER and failure rates are percentages, so we give differences between them in percentage points (pp).

\paragraph{Protocol.}
The same person can appear in several recordings, so training recordings whose voice matches a development or test recording are removed, and cross-validation folds keep same-voice recordings together.
To keep the test honest, all design choices were made by cross-validation on the training recordings, and the selector, the output it selects by and the main comparison (against the UTMOS rule) were fixed before the test sets were scored, so test results could not influence them.
All intervals in the paper are 95\% bootstrap intervals over recordings, resampling same-voice recordings together; we mark a difference ($^\dagger$ better, $^\ddagger$ worse) when its interval excludes zero.
\section{How Much Does the Prompt Matter, and Does Predicted Quality Find Good Prompts?}
\label{sec:sensitivity}
\label{sec:quality}

\begin{table*}[t]
\centering\small
\setlength{\tabcolsep}{4.4pt}
\begin{tabular}{llrrrr|rrrr|rrr}
\toprule
& & \multicolumn{4}{c|}{random prompt} & \multicolumn{4}{c|}{$\Delta$ with probe selection ($\uparrow$ better)} & \multicolumn{3}{c}{ICC} \\
model & prompt & WER & fail. & UTMOS & sim. & WER & UTMOS & sim. & Audiobox & WER & UTMOS & sim. \\
\midrule
\multirow{3}{*}{Qwen3-TTS} & 3\,s & 4.31 & 3.05 & 3.81 & 0.556 & 1.42 & 0.190 & 0.077 & 0.253 & 0.26 & 0.26 & 0.50 \\
 & 5\,s & 4.16 & 2.24 & 3.80 & 0.604 & 1.21 & 0.152 & 0.052 & 0.176 & 0.22 & 0.25 & 0.38 \\
 & 10\,s & 4.01 & 1.64 & 3.78 & 0.650 & 1.10 & 0.097 & 0.023 & 0.104 & 0.15 & 0.20 & 0.22 \\
\midrule
\multirow{3}{*}{Chatterbox} & 3\,s & 2.82 & 0.43 & 3.77 & 0.522 & 0.24 & 0.203 & 0.098 & 0.330 & 0.10 & 0.44 & 0.70 \\
 & 5\,s & 2.77 & 0.49 & 3.79 & 0.573 & 0.20 & 0.148 & 0.064 & 0.219 & 0.07 & 0.34 & 0.57 \\
 & 10\,s & 2.65 & 0.47 & 3.81 & 0.622 & 0.17 & 0.073 & 0.033 & 0.090 & 0.04 & 0.20 & 0.35 \\
\midrule
\multirow{3}{*}{F5-TTS} & 3\,s & 6.24 & 3.18 & 3.26 & 0.538 & 3.28 & 0.253 & 0.099 & 0.402 & 0.35 & 0.43 & 0.70 \\
 & 5\,s & 5.50 & 2.53 & 3.25 & 0.603 & 2.40 & 0.184 & 0.066 & 0.311 & 0.26 & 0.34 & 0.62 \\
 & 10\,s & 4.72 & 1.37 & 3.24 & 0.663 & 1.27 & 0.102 & 0.031 & 0.160 & 0.15 & 0.22 & 0.43 \\
\bottomrule
\end{tabular}
\caption{How much the prompt matters (400 training recordings, candidates of one prompt length). With a random prompt: average over a recording's candidates, WER and failures in \% (failure: WER of at least 50\% or a runaway output of 25\,s or longer). $\Delta$: improvement by probe selection, in pp of WER or in metric units (UTMOS 1--5, Audiobox 1--10, similarity as a cosine). ICC: share of the output variance for one target sentence that is due to the prompt rather than the seed.}
\label{tab:sensitivity}
\end{table*}

We answer both questions on 400 training recordings, on which no selector is evaluated.
As a yardstick for what selection can achieve, we use \emph{probe selection}, which chooses by actually synthesizing from the candidates.
For each target sentence in turn, it picks the candidate whose 12 outputs for the other four sentences score best on average and scores it by its outputs for the held-out sentence, so the choice and its evaluation never share outputs; this is done for each metric separately and averaged over the five sentences.
Probe selection is a practical reference, not a ceiling: judging each prompt from only 12 outputs, it can miss the best one, so a selector can match or exceed it.

\paragraph{The prompt matters, and partly as a stable property of the prompt.}
Choosing among the 3\,s prompts of a recording by probe selection lowers the WER of Qwen3-TTS by about a third relative to a random prompt and that of F5-TTS by half, and it improves naturalness, by every judge, and speaker similarity (Table~\ref{tab:sensitivity}).
These differences are not only sampling noise. For a fixed target sentence, a sizeable share of the variation between outputs is due to which prompt was used rather than to the random seed: the intraclass correlation (ICC; \citealp{shrout1979icc}) in Table~\ref{tab:sensitivity} measures this share, from 0 if the prompt does not matter beyond chance to 1 if it explains all of the variation.
The share is larger for similarity than for intelligibility, so how good a prompt is can in principle be predicted from the prompt.
Chatterbox, which does not read the prompt transcript, is much less sensitive in intelligibility but at least as sensitive in naturalness and similarity.
Longer prompts give more similar voices, but they differ less from each other and fewer of them fit the 30\,s budget, so the room for selection shrinks from 3 to 10\,s.

\paragraph{Predicted quality recovers mainly the naturalness part.}
The common practice is to pick the candidate with the best predicted quality.
We use UTMOS, which targets the naturalness of synthesized and read speech, as the main predictor (the \emph{UTMOS rule}), and DNSMOS, which targets the quality of noise-suppressed speech and filters TTS training data, as the second (the \emph{DNSMOS rule}).
Both improve the naturalness of the outputs, which is what they predict, but recover only part of the intelligibility gain that probe selection shows is available and leave speaker similarity unchanged (all rules in Appendix~\ref{app:rules}).
The UTMOS rule also strongly favours short prompts: among all 19 candidates it picks a 3\,s prompt for 80\% of the recordings, although only about half of the candidates are that short.

\paragraph{Judging the rule with its own predictor flatters it.}
For Qwen3-TTS, the UTMOS rule captures 41\% of probe selection's UTMOS gain but only 18\% of its gain by the held-out Audiobox judge, and for F5-TTS none of it; the other models show the same pattern (Appendix~\ref{app:rules}).
Part of a quality rule's apparent benefit is thus agreement between the selector and the judge, which is why we report every metric and keep Audiobox out of all selectors.
\section{Can Synthesis Outcomes Teach Which Prompt Is Better?}
\label{sec:learned}

\begin{figure*}[t]
\centering
\includegraphics[width=\textwidth]{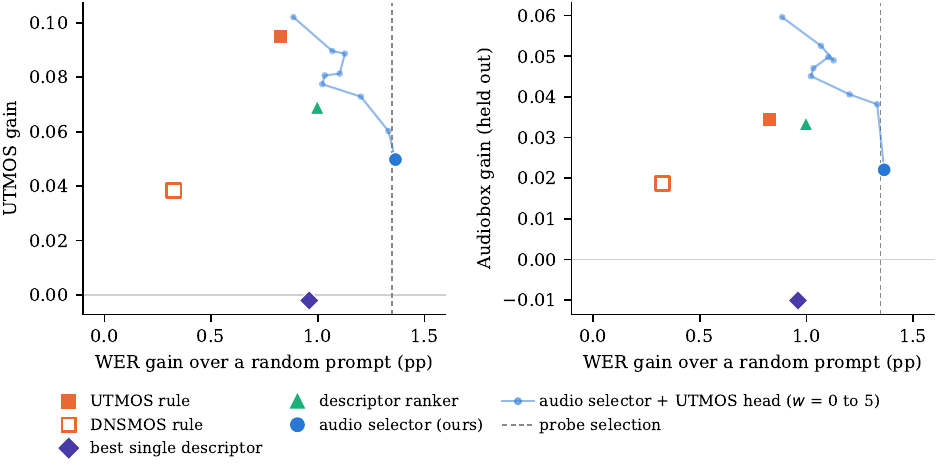}
\caption{Intelligibility gain against naturalness gain on the test (Qwen3-TTS 0.6B), by UTMOS (left) and by the held-out Audiobox judge (right). The curve adds the audio selector's UTMOS head with weight $w$; probe selection's naturalness gains lie above the plotted range.}
\label{fig:tradeoff}
\end{figure*}

\paragraph{From quality rules to learned selection.}
Quality rules rank the candidates by a single predictor.
We extend them in two steps, both learned from the synthesis outcomes of one TTS model at a time.
First, a \emph{descriptor ranker} learns how to weigh 21 descriptors of a prompt: the quality predictors' scores together with speaking rate, pauses and Whisper's confidence in the prompt transcript (all 21 in Appendix~\ref{app:repro}).
Second, the \emph{audio selector} needs no descriptors: it reads the prompt audio through a frozen Whisper-base encoder \citep{radford2023whisper} (21M parameters), and a small trained network (two transformer layers and attention pooling, 1.4M parameters) turns the encoder's output into a score for each prompt, comparing it with the recording's other candidates; we average the scores of five such networks trained from different random initializations, which makes the scores less dependent on one training run \citep{lakshminarayanan2017ensembles}.
Both are trained to rank each recording's candidates in the order of their outcomes, with a listwise softmax loss \citep[ListNet;][]{cao2007listnet}; a pointwise regression loss works as well (\S\ref{sec:analysis}).
They predict several outcomes (the mean WER, the candidate's mean WER rank among its recording's candidates, the failure rate, UTMOS and similarity), and select by the WER rank, since intelligibility is what predicted quality misses and the rank is less sensitive to a few catastrophic outputs (fixed before testing).
All design choices were made by cross-validation on the training recordings (details in Appendix~\ref{app:repro}).

\begin{figure}[t]
\centering
\includegraphics[width=\columnwidth]{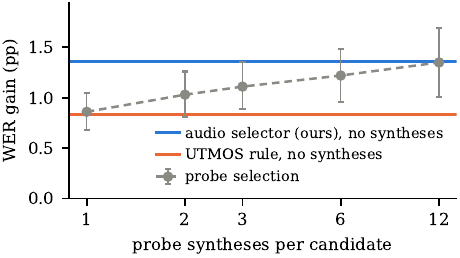}
\caption{What probing costs (Qwen3-TTS 0.6B, test): WER gain over a random prompt of probe selection with 1--12 syntheses per candidate (19--228 per recording), with intervals, against the audio selector and the UTMOS rule, which synthesize nothing.}
\label{fig:probe}
\end{figure}

\paragraph{Results on held-out recordings.}
On the 400 test recordings, the audio selector lowers the WER of Qwen3-TTS 0.6B by 1.36\,pp, 32\% relative to a random prompt (Table~\ref{tab:final}, Figure~\ref{fig:tradeoff}).
This matches probe selection with 12 syntheses per candidate, 228 per recording, while the selector synthesizes nothing (Figure~\ref{fig:probe}); it is 0.54\,pp more than the UTMOS rule (the pre-specified comparison; positive in both test sets and across three training seeds, Appendix~\ref{app:stats}), more than the DNSMOS rule and more than the best single descriptor chosen on the training recordings (Whisper's confidence in the prompt transcript), and it also fails less often than a random prompt.
The descriptor ranker gains less among all 19 candidates, although it nearly matches the audio selector among 3\,s prompts (Appendix~\ref{app:robust}); the audio selector needs none of its inputs, which require running UTMOS, DNSMOS, a speaker model and Whisper decoding on every candidate.

\paragraph{Predicted naturalness is not a reliable judge on its own, and can be traded.}
Selecting for intelligibility gives up part of the UTMOS rule's gain in predicted naturalness (Figure~\ref{fig:tradeoff}), but how large this gap is depends on the judge.
It is large on UTMOS, the predictor the rule selects by; on DNSMOS and the held-out Audiobox judge, the rule's own gain and the gap are both a few hundredths of a MOS point (Table~\ref{tab:final}); and listeners do not prefer the rule's outputs on naturalness (below).
A quality predictor that both selects and judges overstates the naturalness of its own picks, so we do not read the UTMOS gap as a perceived naturalness cost.
Where predicted naturalness matters nonetheless, it can be traded without retraining: the selector already predicts UTMOS, and adding its UTMOS head with weight $w$ to the WER-rank head moves it along the curve in Figure~\ref{fig:tradeoff}.
With $w=0.25$, chosen on the development recordings before testing, it keeps a WER gain of 1.33\,pp, 0.51\,pp more than the UTMOS rule, and matches the rule on DNSMOS and Audiobox; only UTMOS itself still prefers the rule.

\paragraph{The listening test finds no evidence of a naturalness cost.}
In a pairwise listening test on 72 recordings where the picks differ, listeners compared outputs from the selector's and UTMOS rule's prompts for the same target sentence and seed, judging clarity and naturalness (Appendix~\ref{app:human}).
Listeners did not prefer the rule's outputs on naturalness, although UTMOS rates them higher.
On clarity, most pairs were judged similar, as by Whisper: the selector's advantage lies in avoiding less frequent errors that one output per pair rarely samples, so a test of this size cannot resolve it (Appendix~\ref{app:human}).
Where outputs differ in word errors, listeners lean towards the one with fewer errors.

\section{Where Does the Gain Come From?}
\label{sec:analysis}

\paragraph{Not from prompt length, failures, the recognizer or the transcripts.}
The selector's advantage does not come from preferring some prompt length: it beats the UTMOS rule among candidates of each single length, and also with equal numbers of candidates of each length.
Nor does it come only from avoiding failed outputs: with failed outputs excluded, its gain is still about three times the rule's.
Whisper plays three roles: it transcribes the prompts, provides the selector's encoder and judges the outputs, so the selector might learn Whisper's quirks rather than intelligibility.
We therefore re-judged all outputs with a second, unrelated recognizer, wav2vec~2.0 \citep{baevski2020wav2vec}, with a different architecture and training data; by its WER the advantage over the UTMOS rule is even larger (0.71\,pp).
A TTS model that reads the prompt transcript also depends on that transcript, so we resynthesized one of the two test sets with an independent transcript, the YouTube captions shipped with YODAS: the frozen selector, which never saw these transcripts, keeps its margin over the UTMOS rule, nearly unchanged for Qwen3-TTS and slightly smaller for F5-TTS.
Its picks' transcripts are about as reliable as the rule's, and among prompts whose Whisper transcript and YouTube caption agree it stays ahead of the rule.
Nor does selecting for intelligibility reduce pitch variability more than the quality rule: it drops slightly under both, and measured against the speaker's own reading of each target sentence, the selector's outputs match its pitch variability as well as a random prompt's and its tempo slightly better (Appendix~\ref{app:robust}).
Details of all these checks are in Appendix~\ref{app:robust}.

\begin{table}[tb]
\centering\small
\setlength{\tabcolsep}{4.5pt}
\begin{tabular}{lrrr}
\toprule
variant (cross-validation) & WER & $\Delta$ & \shortstack{descr.\\ranker} \\
\midrule
final recipe & 1.25 &  & 0.90 \\
\quad pointwise loss & 1.18 & -0.07 & 0.90 \\
\quad labels: prompt's UTMOS & 0.63 & -0.62 & 0.55 \\
\quad Whisper-large-v3 encoder & 1.22 & -0.03 & 0.90 \\
\quad CLAP encoder & 0.42 & -0.82 & 0.90 \\
\bottomrule
\end{tabular}
\caption{Selected ablations by cross-validation on the training recordings (mean of three seeds): WER gain over a random prompt (pp), its change from the final recipe, and the descriptor ranker trained with the same labels on the same folds. All ablations in Appendix~\ref{app:robust}.}
\label{tab:ablations}
\end{table}

\begin{figure*}[t]
\centering
\includegraphics[width=0.78\textwidth]{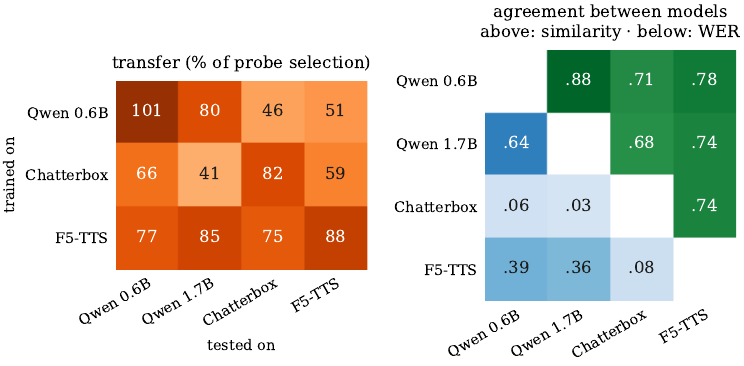}
\caption{Left: transfer between TTS models, as the WER gain of a selector trained on one model (rows) for another (columns), in percent of the second model's probe-selection gain. Right: do two models prefer the same prompts? Within-recording Spearman correlation between the candidates' speaker similarity (above the diagonal) and WER (below) under two models, averaged over the 400 test recordings and corrected for how consistently each model ranks the candidates between two halves of its own outputs (split-half reliability; uncorrected values in Appendix~\ref{app:models}).}
\label{fig:agreement}
\end{figure*}

\paragraph{What matters is the supervision by synthesis outcomes.}
Varying one choice at a time by cross-validation (Table~\ref{tab:ablations}), the architecture hardly matters: a pointwise regression loss instead of the listwise one, a single network instead of five, or attention across a recording's candidates all stay within the spread between training seeds (about 0.1\,pp).
We found that speech encoders work better than a general-audio encoder (CLAP, \citealp{wu2023clap}), which gains little, and that a small one suffices: Whisper-base does as well as Whisper-large-v3 with a thirtieth of its parameters, and WavLM \citep{chen2022wavlm}, which is not trained for recognition, does as well.
What matters is the training signal: the same model trained to reproduce each candidate's own predicted quality instead of its synthesis outcomes gains about half as much.
A good prompt is defined by what the TTS model makes of it, and that has to be measured.
\section{Do TTS Models Agree on What Makes a Good Prompt?}
\label{sec:models}

\paragraph{Trained for each model, the selector matches or beats the strongest simple baseline.}
We train the same recipe on the outcomes of F5-TTS and of Chatterbox and compare each selector with the quality rules and with the best single descriptor chosen on that model's training recordings (Figure~\ref{fig:models}).
Which simple baseline is strongest depends on the model, and the selector is ahead of it for three of the four models: clearly for F5-TTS, whose WER it lowers by 43\% relative to a random prompt, 1.45\,pp more than its strongest baseline (the DNSMOS rule), and for Qwen3-TTS 1.7B, and narrowly for Qwen3-TTS 0.6B, against Whisper's confidence in the prompt transcript; for Chatterbox, which has little room for selection, the two are level.
On the other metrics it stays level with the UTMOS rule or better, except UTMOS itself (Appendix~\ref{app:models}).

\paragraph{Models largely agree on prompts for speaker similarity.}
A selector trained on one model's outcomes transfers to the other models without retraining, to varying degrees (Figure~\ref{fig:agreement}, left): the selector trained on Qwen3-TTS 0.6B gains 80\% of probe selection's gain for the 1.7B model, where the UTMOS rule barely helps, and 51\% for F5-TTS, and the selector trained on F5-TTS transfers best of all, with 77--85\% for the Qwen models.
Speaker similarity is where the models agree most: which candidate yields the highest measured speaker similarity is largely the same for all four models (Figure~\ref{fig:agreement}, right, above the diagonal; corrected correlations 0.68--0.88), a property of the prompt rather than of the model.

\begin{table}[tb]
\centering\small
\setlength{\tabcolsep}{2.6pt}
\begin{tabular}{lrrrr}
\toprule
prompt property & \shortstack{Qwen\\0.6B} & \shortstack{Qwen\\1.7B} & \shortstack{Chat-\\terbox} & \shortstack{F5-\\TTS} \\
\midrule
UTMOS & +0.04$^\dagger$ & +0.05$^\dagger$ & +0.05$^\dagger$ & +0.09$^\dagger$ \\
typical voice & -0.00 & -0.01 & +0.05$^\dagger$ & +0.07$^\dagger$ \\
ends in silence & +0.16$^\dagger$ & +0.16$^\dagger$ & +0.01 & +0.11$^\dagger$ \\
transcript confidence & +0.10$^\dagger$ & +0.10$^\dagger$ & +0.03$^\dagger$ & +0.10$^\dagger$ \\
\bottomrule
\end{tabular}
\caption{Prompt properties and intelligibility, per TTS model: within-recording Spearman correlation between a property of the prompt and the intelligibility of its outputs (fewer word errors = higher), averaged over the 400 test recordings. Typical voice: speaker embedding close to the recording's mean; ends in silence: length of the trailing silence; transcript confidence: Whisper's mean token log-probability. $^\dagger$: interval excludes zero. All 21 descriptors in Appendix~\ref{app:what}.}
\label{tab:descr}
\end{table}

\paragraph{The rest depends on how the model uses the prompt.}
For each model with training outcomes, the selector trained on that model is the best one for it (Figure~\ref{fig:agreement}, left), and which candidates give intelligible speech is largely model-specific (right, below the diagonal).
The two Qwen models agree most (0.64), Qwen3-TTS and F5-TTS, which both read the prompt transcript, agree in part (0.39), and Chatterbox, which does not, agrees with neither (at most 0.08).
The prompt properties show the same split (Table~\ref{tab:descr}): for the models that read the transcript, prompts that end in silence rather than in speech, and prompts whose transcript Whisper is confident about, give more intelligible outputs, while neither matters for Chatterbox; how typical the prompt's voice is of the recording matters for Chatterbox and F5-TTS, and predicted quality only a little for any model.
Transfer follows this agreement only loosely (rank correlation 0.67 over the 9 pairs of models), because a selector also carries the part that all models share.
No single descriptor correlates with a prompt's intelligibility by more than about 0.17 within a recording, for any model, so the model-specific part has to be learned from that model's outcomes.
\section{What If Prompts Differ in Noise?}
\label{sec:noise}

On YouTube recordings the candidates of a recording share their recording conditions, and predicted quality recovers little of what makes a good prompt.
Prompts may also differ mainly in recording quality, e.g., when recorded in different places or with different devices.
To study this under controlled conditions, we start with clean read speech (LibriSpeech \citep{panayotov2015librispeech}) and add noise of random type and level to some candidates, retaining knowledge of which version is clean (Appendix~\ref{app:noise}). Outputs come from Qwen3-TTS 1.7B, while the selector is the frozen audio selector trained on YouTube outcomes from Qwen3-TTS 0.6B, which has seen no added noise.

\begin{table}[tb]
\centering\small
\setlength{\tabcolsep}{3.5pt}
\begin{tabular}{lrrrrr}
\toprule
selector & WER & fail. & UTMOS & sim. & clean \\
\midrule
UTMOS rule & 0.21 & -0.05 & \textbf{0.550} & \textbf{0.113} & \textbf{79\%} \\
audio selector & \textbf{0.63} & \textbf{0.26} & 0.349 & 0.070 & 31\% \\
\quad + UTMOS head & \textbf{0.66} & \textbf{0.26} & 0.415 & 0.080 & 39\% \\
probe selection & 0.42 & 0.18 & 0.570 & 0.165 & \\
\bottomrule
\end{tabular}
\caption{Candidates differing in noise (LibriSpeech, Qwen3-TTS 1.7B, all lengths in one selection): gain over a random prompt, WER and failures in pp, and the share of picks that are the clean version of a prompt (by chance: 14\%, one of seven conditions); the UTMOS head is added with $w=0.25$. The selectors were trained on YouTube outcomes of Qwen3-TTS 0.6B.}
\label{tab:noise}
\end{table}

Here the quality rule does what it is meant to do: it picks the clean version of a prompt 79\% of the time and gains the most naturalness and similarity (Table~\ref{tab:noise}).
It does not find intelligible prompts, though: the audio selector lowers WER by 0.42\,pp more and fails less often, while it picks the clean version only 31\% of the time and gives up UTMOS and similarity.
Adding the UTMOS head moves the selector towards the rule on naturalness and keeps its intelligibility.
Probe selection, ranking candidates by a dozen outputs each, gains less WER than the selector at this low error rate.
When prompts differ in noise, predicted quality is the right tool for naturalness and similarity, but intelligibility depends on other properties of the prompt, which the selector has learned from the outcomes.
\section{Conclusion}
\label{sec:conclusion}

We asked what makes a good audio prompt for zero-shot TTS, judging each prompt by the speech a model produces from it. The answer lies in both the prompt and the model. TTS models largely agree on which prompt yields the highest measured speaker similarity, but much less on which yields intelligible speech. Choosing by predicted audio quality, the common practice, improves mainly naturalness, looks better than it is when the same predictor also judges the result, and finds the clean prompt rather than the intelligible one. Learning from synthesis outcomes recovers what predicted quality misses. A small network that hears only the candidates' audio comes close to choosing by synthesizing a dozen test sentences from every candidate, and works best when trained on the model it selects for.

These findings have three practical implications. For users of zero-shot TTS, a prompt is worth choosing by what the model makes of it rather than by how clean it sounds, and a small selector makes this cheap. For benchmarks, pairing each sentence with one fixed prompt partly mixes up the quality of the model with the luck of the prompt; reporting how much outputs vary across prompts would separate the two. For model developers, the fact that sensitivity to the prompt differs between models suggests that it can be reduced.

Our selector works at inference time, for a fixed model. The same outcomes could also guide training, for example which prompts a zero-shot model is trained with, or an objective that makes it less sensitive to the prompt. Collecting outcomes could be made cheaper, with fewer outputs per prompt, as our ablations suggest, or by sharing them across models. A selector could also take the target sentence into account when it is known in advance.
\section*{Limitations}

We study four TTS models from three families, on English YouTube recordings and on LibriSpeech with added noise; other languages, for which zero-shot models are often trained on weakly supervised in-the-wild speech \citep{he2024emilia,turetzky2024hebdb,sankar2024indicvoicesr,roth2024hebtts}, and other recording conditions are open; the method itself uses no language-specific component apart from Whisper, which is multilingual.
The best selector for a model is trained on that model's outcomes, which costs one labelling run of a few hundred thousand syntheses per model (Appendix~\ref{app:repro}), paid once per model rather than per voice; how far fewer labels or a selector trained for another model can replace it is only partly explored (Appendix~\ref{app:robust}, \S\ref{sec:models}).
A prompt's outcome averages 15 outputs and WER is heavy-tailed, so outcomes are noisy, and probe selection, estimated from 12 outputs per candidate, is not an upper bound.
Apart from the listening test, all measurements are automatic; the listening test, with 7 listeners and one output per pair, finds no evidence of a naturalness cost but is too small to resolve the selector's clarity advantage (Appendix~\ref{app:human}).
Beyond pitch variability and tempo (Appendix~\ref{app:robust}), we do not evaluate prosody, e.g.\ emphasis, expressiveness and context-appropriate stress, which dedicated benchmarks measure \citep{deseyssel2024emphassess,turetzky26_interspeech,manku2025emergenttts}; how the prompt affects them is open.
The test has 400 recordings, and the selector's margin over the UTMOS rule varies between training seeds (Appendix~\ref{app:stats}).
We fix a 30\,s budget and prompts of 3--10\,s; longer prompts and concatenations of several prompts are not studied.
\section*{Ethical Considerations}

Voice cloning can be misused for impersonation and fraud.
Selecting a better prompt makes clones more intelligible rather than more similar to the speaker (our selectors improve speaker similarity little or not at all), but it lowers the effort of producing usable clones.
We use public CC-BY YouTube speech (YODAS) and LibriSpeech for research only, distribute no voice models, speaker audio or synthesized speech of the speakers, and report aggregate results.
A Creative Commons license permits reuse of the recordings, but it is not the speakers' consent to having their voices cloned, so all synthesized speech stays internal, used only for evaluation and the listening test.

\bibliography{refs}

@article{wang2023valle,
  title   = {Neural Codec Language Models are Zero-Shot Text to Speech Synthesizers},
  author  = {Wang, Chengyi and Chen, Sanyuan and Wu, Yu and Zhang, Ziqiang and Zhou, Long and Liu, Shujie and Chen, Zhuo and Liu, Yanqing and Wang, Huaming and Li, Jinyu and He, Lei and Zhao, Sheng and Wei, Furu},
  journal = {arXiv preprint arXiv:2301.02111},
  year    = {2023}
}

@article{anastassiou2024seedtts,
  title   = {{Seed-TTS}: A Family of High-Quality Versatile Speech Generation Models},
  author  = {Anastassiou, Philip and others},
  journal = {arXiv preprint arXiv:2406.02430},
  year    = {2024}
}

@inproceedings{chen2024f5tts,
  title     = {{F5-TTS}: A Fairytaler that Fakes Fluent and Faithful Speech with Flow Matching},
  author    = {Chen, Yushen and Niu, Zhikang and Ma, Ziyang and Deng, Keqi and Wang, Chunhui and Zhao, Jian and Yu, Kai and Chen, Xie},
  booktitle = {Proceedings of ACL},
  pages     = {6255--6271},
  doi       = {10.18653/v1/2025.acl-long.313},
  year      = {2025}
}

@article{du2024cosyvoice2,
  title   = {{CosyVoice} 2: Scalable Streaming Speech Synthesis with Large Language Models},
  author  = {Du, Zhihao and Wang, Yuxuan and Chen, Qian and Shi, Xian and Lv, Xiang and Zhao, Tianyu and Gao, Zhifu and Yang, Yexin and Gao, Changfeng and Wang, Hui and Yu, Fan and Liu, Huadai and Sheng, Zhengyan and Gu, Yue and Deng, Chong and Wang, Wen and Zhang, Shiliang and Yan, Zhijie and Zhou, Jingren},
  journal = {arXiv preprint arXiv:2412.10117},
  year    = {2024}
}

@inproceedings{wang2025maskgct,
  title     = {{MaskGCT}: Zero-Shot Text-to-Speech with Masked Generative Codec Transformer},
  author    = {Wang, Yuancheng and Zhan, Haoyue and Liu, Liwei and Zeng, Ruihong and Guo, Haotian and Zheng, Jiachen and Zhang, Qiang and Zhang, Xueyao and Zhang, Shunsi and Wu, Zhizheng},
  booktitle = {Proceedings of ICLR},
  year      = {2025}
}

@inproceedings{meng2025melle,
  title     = {Autoregressive Speech Synthesis without Vector Quantization},
  author    = {Meng, Lingwei and Zhou, Long and Liu, Shujie and Chen, Sanyuan and Han, Bing and Hu, Shujie and Liu, Yanqing and Li, Jinyu and Zhao, Sheng and Wu, Xixin and Meng, Helen M. and Wei, Furu},
  booktitle = {Proceedings of ACL},
  year      = {2025}
}

@inproceedings{turetzky2025salad,
  title     = {Speech Synthesis From Continuous Features Using Per-Token Latent Diffusion},
  author    = {Turetzky, Arnon and Dekel, Avihu and Shabtay, Nimrod and Shechtman, Slava and Haws, David and Aronowitz, Hagai and Hoory, Ron and Adi, Yossi},
  booktitle = {Proceedings of ASRU},
  pages     = {1--8},
  doi       = {10.1109/ASRU65441.2025.11434698},
  year      = {2025}
}

@article{qwen3tts,
  title   = {{Qwen3-TTS} Technical Report},
  author  = {Hu, Hangrui and Zhu, Xinfa and He, Ting and Guo, Dake and Zhang, Bin and Wang, Xiong and Guo, Zhifang and Jiang, Ziyue and Hao, Hongkun and Guo, Zishan and Zhang, Xinyu and Zhang, Pei and Yang, Baosong and Xu, Jin and Zhou, Jingren and Lin, Junyang},
  journal = {arXiv preprint arXiv:2601.15621},
  year    = {2026}
}

@misc{chatterbox,
  title        = {Chatterbox {TTS}},
  author       = {{Resemble AI}},
  year         = {2025},
  howpublished = {\url{https://github.com/resemble-ai/chatterbox}}
}

@article{betker2023tortoise,
  title   = {Better Speech Synthesis Through Scaling},
  author  = {Betker, James},
  journal = {arXiv preprint arXiv:2305.07243},
  year    = {2023}
}

@inproceedings{gong2021multiref,
  title     = {Using Multiple Reference Audios and Style Embedding Constraints for Speech Synthesis},
  author    = {Gong, Cheng and Wang, Longbiao and Ling, Zhenhua and Zhang, Ju and Dang, Jianwu},
  booktitle = {Proceedings of ICASSP},
  year      = {2022}
}

@article{raon2026opentts,
  title   = {Raon-{OpenTTS}: Open Models and Data for Robust Text-to-Speech},
  author  = {Kim, Semin and Chung, Seungjun and Moon, Taehong and Lee, Sangheon and Ahn, Minyoung and Lee, Keon and Kim, Nam Soo and Cho, Jaewoong and Schmidt, Ludwig and Lee, Kangwook and Park, Dongmin},
  journal = {arXiv preprint arXiv:2605.20830},
  year    = {2026}
}

@inproceedings{luo2025autostyle,
  title     = {{AutoStyle-TTS}: Retrieval-Augmented Generation based Automatic Style Matching Text-to-Speech Synthesis},
  author    = {Luo, Dan and Ma, Chengyuan and Li, Weiqin and Wang, Jun and Chen, Wei and Wu, Zhiyong},
  booktitle = {Proceedings of ICME},
  year      = {2025}
}

@article{wang2024expressive,
  title   = {Expressive Prompting: Improving Emotion Intensity and Speaker Consistency in Zero-Shot {TTS}},
  author  = {Wang, Haoyu and Qiang, Chunyu and Wang, Tianrui and Gong, Cheng and Jiang, Yu and Lu, Yuheng and Zhang, Chen and Wang, Longbiao and Dang, Jianwu},
  journal = {arXiv preprint arXiv:2409.18512},
  year    = {2024}
}

@article{zhu2025m3pdb,
  title   = {{M$^3$PDB}: A Multimodal, Multi-Label, Multilingual Prompt Database for Speech Generation},
  author  = {Zhu, Boyu and Gong, Cheng and Wu, Muyang and Jing, Ruihao and Liu, Fan and Zhang, Xiaolei and Zhang, Chi and Li, Xuelong},
  journal = {arXiv preprint arXiv:2508.09702},
  year    = {2025}
}

@inproceedings{giraldo2026bsc,
  title     = {Zero-Shot {TTS} with Enhanced Audio Prompts: {BSC} Submission for the 2026 {WildSpoof} Challenge {TTS} Track},
  author    = {Giraldo, Jose and Peir{\'o}-Lilja, Alex and Zevallos, Rodolfo and Espa{\~n}a-Bonet, Cristina},
  booktitle = {Proceedings of ICASSP},
  year      = {2026}
}

@article{sun2026dps,
  title   = {Natural Yet Challenging to Detect: Robust In-the-Wild {TTS} through {EMA} and Dual-Scoring Prompt Selection -- Submission for {WildSpoof} 2026 {TTS} Track},
  author  = {Sun, Renhe and Zhou, Jiayi and He, Haolin and Feng, Yueying and Liu, Jian},
  journal = {arXiv preprint arXiv:2605.23859},
  year    = {2026}
}

@inproceedings{panayotov2015librispeech,
  title={Librispeech: An ASR corpus based on public domain audio books},
  author={Panayotov, Vassil and Chen, Guo-Guo and Povey, Daniel and Khudanpur, Sanjeev},
  booktitle={2015 IEEE International Conference on Acoustics, Speech and Signal Processing (ICASSP)},
  pages={5206--5210},
  year={2015},
  organization={IEEE}
}

@inproceedings{li2023yodas,
  title     = {{YODAS}: Youtube-Oriented Dataset for Audio and Speech},
  author    = {Li, Xinjian and Takamichi, Shinnosuke and Saeki, Takaaki and Chen, William and Shiota, Sayaka and Watanabe, Shinji},
  booktitle = {Proceedings of ASRU},
  year      = {2023}
}

@inproceedings{he2024emilia,
  title     = {Emilia: An Extensive, Multilingual, and Diverse Speech Dataset for Large-Scale Speech Generation},
  author    = {He, Haorui and Shang, Zengqiang and Wang, Chaoren and Li, Xuyuan and Gu, Yicheng and Hua, Hua and Liu, Liwei and Yang, Chen and Li, Jiaqi and Shi, Peiyang and Wang, Yuancheng and Chen, Kai and Zhang, Pengyuan and Wu, Zhizheng},
  booktitle = {Proceedings of SLT},
  year      = {2024}
}

@inproceedings{turetzky2024hebdb,
  title     = {{HebDB}: a Weakly Supervised Dataset for {Hebrew} Speech Processing},
  author    = {Turetzky, Arnon and Tal, Or and Segal-Feldman, Yael and Dissen, Yehoshua and Zeldes, Ella and Roth, Amit and Cohen, Eyal and Shrem, Yosi and Chernyak, Bronya R. and Seleznova, Olga and Keshet, Joseph and Adi, Yossi},
  booktitle = {Proceedings of Interspeech},
  pages     = {1360--1364},
  year      = {2024}
}

@inproceedings{sankar2024indicvoicesr,
  title     = {{IndicVoices-R}: Unlocking a Massive Multilingual Multi-speaker Speech Corpus for Scaling {Indian} {TTS}},
  author    = {Sankar, Ashwin and Anand, Srija and Varadhan, Praveen Srinivasa and Thomas, Sherry and Singal, Mehak and Kumar, Shridhar and Mehendale, Deovrat and Krishana, Aditi and Raju, Giri and Khapra, Mitesh},
  booktitle = {Proceedings of NeurIPS (Datasets and Benchmarks Track)},
  year      = {2024}
}

@inproceedings{roth2024hebtts,
  title     = {A Language Modeling Approach to Diacritic-Free {Hebrew} {TTS}},
  author    = {Roth, Amit and Turetzky, Arnon and Adi, Yossi},
  booktitle = {Proceedings of Interspeech},
  pages     = {2775--2779},
  year      = {2024}
}

@inproceedings{thiemann2013demand,
  title     = {The Diverse Environments Multi-channel Acoustic Noise Database ({DEMAND}): A Database of Multichannel Environmental Noise Recordings},
  author    = {Thiemann, Joachim and Ito, Nobutaka and Vincent, Emmanuel},
  booktitle = {Proceedings of Meetings on Acoustics (ICA 2013)},
  volume    = {19},
  year      = {2013}
}

@inproceedings{saeki2022utmos,
  title     = {{UTMOS}: {UTokyo-SaruLab} System for {VoiceMOS} Challenge 2022},
  author    = {Saeki, Takaaki and Xin, Detai and Nakata, Wataru and Koriyama, Tomoki and Takamichi, Shinnosuke and Saruwatari, Hiroshi},
  booktitle = {Proceedings of Interspeech},
  year      = {2022}
}

@inproceedings{reddy2022dnsmos,
  title     = {{DNSMOS P.835}: A Non-Intrusive Perceptual Objective Speech Quality Metric to Evaluate Noise Suppressors},
  author    = {Reddy, Chandan K. A. and Gopal, Vishak and Cutler, Ross},
  booktitle = {Proceedings of ICASSP},
  year      = {2022}
}

@article{tjandra2025audiobox,
  title   = {Meta Audiobox Aesthetics: Unified Automatic Quality Assessment for Speech, Music, and Sound},
  author  = {Tjandra, Andros and Wu, Yi-Chiao and Guo, Baishan and Hoffman, John and Ellis, Brian and Vyas, Apoorv and Shi, Bowen and Chen, Sanyuan and Le, Matt and Zacharov, Nick and Wood, Carleigh and Lee, Ann and Hsu, Wei-Ning},
  journal = {arXiv preprint arXiv:2502.05139},
  year    = {2025}
}

@inproceedings{radford2023whisper,
  title     = {Robust Speech Recognition via Large-Scale Weak Supervision},
  author    = {Radford, Alec and Kim, Jong Wook and Xu, Tao and Brockman, Greg and McLeavey, Christine and Sutskever, Ilya},
  booktitle = {Proceedings of ICML},
  year      = {2023}
}

@inproceedings{baevski2020wav2vec,
  title     = {wav2vec 2.0: A Framework for Self-Supervised Learning of Speech Representations},
  author    = {Baevski, Alexei and Zhou, Henry and Mohamed, Abdelrahman and Auli, Michael},
  booktitle = {Proceedings of NeurIPS},
  year      = {2020}
}

@article{chen2022wavlm,
  title   = {{WavLM}: Large-Scale Self-Supervised Pre-Training for Full Stack Speech Processing},
  author  = {Chen, Sanyuan and Wang, Chengyi and Chen, Zhengyang and Wu, Yu and Liu, Shujie and Chen, Zhuo and Li, Jinyu and Kanda, Naoyuki and Yoshioka, Takuya and Xiao, Xiong and others},
  journal = {IEEE Journal of Selected Topics in Signal Processing},
  year    = {2022}
}

@inproceedings{wu2023clap,
  title     = {Large-Scale Contrastive Language-Audio Pretraining with Feature Fusion and Keyword-to-Caption Augmentation},
  author    = {Wu, Yusong and Chen, Ke and Zhang, Tianyu and Hui, Yuchen and Berg-Kirkpatrick, Taylor and Dubnov, Shlomo},
  booktitle = {Proceedings of ICASSP},
  year      = {2023}
}

@inproceedings{desplanques2020ecapa,
  title     = {{ECAPA-TDNN}: Emphasized Channel Attention, Propagation and Aggregation in {TDNN} Based Speaker Verification},
  author    = {Desplanques, Brecht and Thienpondt, Jenthe and Demuynck, Kris},
  booktitle = {Proceedings of Interspeech},
  year      = {2020}
}

@inproceedings{zhao2021calibrate,
  title     = {Calibrate Before Use: Improving Few-Shot Performance of Language Models},
  author    = {Zhao, Tony Z. and Wallace, Eric and Feng, Shi and Klein, Dan and Singh, Sameer},
  booktitle = {Proceedings of ICML},
  year      = {2021}
}

@inproceedings{lu2022fantastically,
  title     = {Fantastically Ordered Prompts and Where to Find Them: Overcoming Few-Shot Prompt Order Sensitivity},
  author    = {Lu, Yao and Bartolo, Max and Moore, Alastair and Riedel, Sebastian and Stenetorp, Pontus},
  booktitle = {Proceedings of ACL},
  year      = {2022}
}

@inproceedings{liu2022makes,
  title     = {What Makes Good In-Context Examples for {GPT}-3?},
  author    = {Liu, Jiachang and Shen, Dinghan and Zhang, Yizhe and Dolan, Bill and Carin, Lawrence and Chen, Weizhu},
  booktitle = {Proceedings of DeeLIO},
  year      = {2022}
}

@inproceedings{rubin2022learning,
  title     = {Learning To Retrieve Prompts for In-Context Learning},
  author    = {Rubin, Ohad and Herzig, Jonathan and Berant, Jonathan},
  booktitle = {Proceedings of NAACL},
  year      = {2022}
}

@article{ticl2025,
  title   = {{TICL}: Text-Embedding {KNN} For Speech In-Context Learning Unlocks Speech Recognition Abilities of Large Multimodal Models},
  author  = {Zheng, Haolong and Yegorova, Yekaterina and Hasegawa-Johnson, Mark},
  journal = {arXiv preprint arXiv:2509.13395},
  year    = {2025}
}

@inproceedings{cao2007listnet,
  title     = {Learning to Rank: From Pairwise Approach to Listwise Approach},
  author    = {Cao, Zhe and Qin, Tao and Liu, Tie-Yan and Tsai, Ming-Feng and Li, Hang},
  booktitle = {Proceedings of ICML},
  year      = {2007}
}

@inproceedings{lakshminarayanan2017ensembles,
  title     = {Simple and Scalable Predictive Uncertainty Estimation using Deep Ensembles},
  author    = {Lakshminarayanan, Balaji and Pritzel, Alexander and Blundell, Charles},
  booktitle = {Proceedings of NeurIPS},
  year      = {2017}
}

@article{shrout1979icc,
  title   = {Intraclass Correlations: Uses in Assessing Rater Reliability},
  author  = {Shrout, Patrick E. and Fleiss, Joseph L.},
  journal = {Psychological Bulletin},
  volume  = {86},
  number  = {2},
  pages   = {420--428},
  year    = {1979}
}

@inproceedings{deseyssel2024emphassess,
  title     = {{EmphAssess}: a Prosodic Benchmark on Assessing Emphasis Transfer in Speech-to-Speech Models},
  author    = {de Seyssel, Maureen and D'Avirro, Antony and Williams, Adina and Dupoux, Emmanuel},
  booktitle = {Proceedings of EMNLP},
  year      = {2024}
}

@inproceedings{turetzky26_interspeech,
  title     = {Knowing What to Stress: A Discourse-Conditioned Text-to-Speech Benchmark},
  author    = {Turetzky, Arnon and Dekel, Avihu and Aronowitz, Hagai and Hoory, Ron and Adi, Yossi},
  booktitle = {Proceedings of Interspeech},
  pages     = {2932--2936},
  doi       = {10.21437/Interspeech.2026-2743},
  year      = {2026}
}

@inproceedings{manku2025emergenttts,
  title     = {{EmergentTTS-Eval}: Evaluating {TTS} Models on Complex Prosodic, Expressiveness, and Linguistic Challenges Using Model-as-a-Judge},
  author    = {Manku, Ruskin Raj and Tang, Yuzhi and Shi, Xingjian and Li, Mu and Smola, Alex},
  booktitle = {Proceedings of NeurIPS (Datasets and Benchmarks Track)},
  year      = {2025}
}

\appendix
\renewcommand{\dbltopfraction}{0.85}\renewcommand{\dblfloatpagefraction}{0.6}\setcounter{dbltopnumber}{3}

\section{Data, Models and Training}
\label{app:repro}

\begin{table*}[tp]
\centering\small
\begin{tabular}{lrcrl}
\toprule
set & recordings & prompts (3/5/10\,s) & outputs per model & TTS models \\
\midrule
train & 400 & 3993 / 2389 / 1170 & 113,280 & Qwen3-TTS, Chatterbox, F5-TTS \\
train & 544 & 5429 / 3249 / 1588 & 153,990 & Qwen3-TTS, Chatterbox, F5-TTS \\
dev & 200 & 2000 / 1194 / 583 & 56,655 & Qwen3-TTS, Chatterbox \\
test & 200 & 1994 / 1194 / 591 & 56,685 & Qwen3-TTS, Chatterbox, F5-TTS \\
test & 200 & 1997 / 1194 / 588 & 56,685 & Qwen3-TTS, Chatterbox, F5-TTS \\
\bottomrule
\end{tabular}
\caption{Recording sets (disjoint YODAS shards): two training sets, a development set and two test sets. Outputs per model: every candidate prompt $\times$ 5 target sentences $\times$ 3 seeds.}
\label{tab:data}
\end{table*}

\paragraph{Data processing.}
YODAS English shards 0000--0002 (development), 0003--0006 and 0009--0012 (training), 0007--0008 and 0013--0014 (test).
Videos of 3--30 minutes with at least 7\,kHz bandwidth; caption lines are merged into 3--20\,s utterances; the main speaker keeps the utterances whose speaker embedding has a cosine of at least 0.88 to that of the video's most typical utterance (the one most similar to all others), and these must make up at least 70\% of the utterances; the spoken language is English by majority vote of Whisper large-v3 language identification on three clips.
Each recording is scaled to a peak amplitude of 0.99.
The 30\,s stretch is drawn so that the main speaker covers at least 80\% of it and of every prompt, so that prompts rarely contain another voice; target sentences and anchors are taken at least 2\,s away from it.
Prompt transcripts with a zlib compression ratio above 2.4 or more than 6 words per second are decoded again with Whisper's temperature fallback (temperatures 0--1, compression threshold 1.35, log-probability threshold $-1$); target sentences that still fail the same test are excluded from all outcomes.

\paragraph{TTS settings.}
Qwen3-TTS (12\,Hz Base, in-context mode with prompt audio and transcript): temperature 0.9, top-$k$ 50, repetition penalty 1.05, batches of 16.
Chatterbox (prompt audio only): temperature 0.8, top-$p$ 1.0, min-$p$ 0.05, repetition penalty 1.2, classifier-free guidance weight 0.5, at most 750 tokens.
F5-TTS v1 Base (prompt audio and transcript): 32 flow steps, guidance strength 2, sway sampling $-1$; the output length follows the prompt's speaking rate, and when this exceeds the model's maximum length (a transcript that is short for its prompt) the output is capped at 30\,s, which counts as a failure.
Every (prompt, sentence, seed) output has its own derived seed, so outputs are reproducible.

\paragraph{Outcomes.}
WER: Whisper large-v3, English, both texts normalized with Whisper's English normalizer followed by removal of tokens without letters or digits, capped at 100\% per output; failure: raw WER of at least 50\% or an output of at least 25\,s.
UTMOS: UTMOS22 strong (SpeechMOS v1.2.0); DNSMOS: P.835 overall; AES: Audiobox Aesthetics production quality; similarity: SpeechBrain ECAPA-TDNN (VoxCeleb), cosine between the output and the mean embedding of three anchor utterances.

\paragraph{Audio selector.}
Encoder: frozen Whisper-base encoder, last layer, averaged to 40\,ms frames; each input dimension standardized with training statistics.
Prompt model: linear projection to 256 dimensions, sinusoidal positions, two transformer layers (4 heads, feed-forward 512, dropout 0.1), attention pooling with one learned query.
Comparison with the recording: the prompt vector concatenated with its difference from the mean prompt vector of the recording; scoring head: 512$\rightarrow$256$\rightarrow$6 (GELU).
Heads: mean WER, WER rank within the recording, failure rate, UTMOS, DNSMOS, ECAPA similarity; selection by the WER-rank head.
Training: ListNet (temperature 1) over all candidates of a recording, AdamW (learning rate $10^{-4}$, weight decay $10^{-2}$), 16 recordings per batch, 10 epochs, bf16 training and fp32 inference; five models with different initializations, scores averaged.
Training on the 782 training recordings (14.8k prompts) takes a few minutes on one GPU.

\paragraph{Descriptor ranker.}
\label{app:what}
Linear listwise ranker with the same loss and labels on the 21 descriptors of Table~\ref{tab:descriptors}, each also z-scored within the recording (typical voice: speaker-embedding cosine to the recording's mean prompt; transcript log-probabilities: Whisper's token log-probabilities of the prompt transcript).
AdamW (learning rate $10^{-3}$, weight decay $10^{-4}$), 32 recordings per batch, 100 epochs.

\paragraph{Splits, leakage and model selection.}
Training recordings whose voice matches a development or test recording (symmetric ECAPA cosine between one recording's anchors and another's target sentences above 0.5) are removed (162 of 944).
Cross-validation: 5 folds over the remaining training recordings, keeping connected components of the same-voice graph in one fold; all hyperparameters were chosen this way.
The same components are the clusters that the bootstrap resamples together.

\paragraph{Compute.}
Labelling costs about 1--2.5\,s of GPU time per output (NVIDIA L40S), i.e.\ about 100--270 GPU hours per model for the training sets.
The recordings are available from YODAS.

\begin{table*}[t!]
\centering\small
\begin{tabular}{lrrrr}
\toprule
descriptor & Qwen3-TTS 0.6B & Qwen3-TTS 1.7B & Chatterbox & F5-TTS \\
\midrule
UTMOS & +0.04$^\dagger$ & +0.05$^\dagger$ & +0.05$^\dagger$ & +0.09$^\dagger$ \\
DNSMOS overall & +0.02 & -0.01 & +0.05$^\dagger$ & +0.09$^\dagger$ \\
DNSMOS signal & +0.01 & +0.01 & +0.05$^\dagger$ & +0.08$^\dagger$ \\
DNSMOS background & +0.01 & -0.01 & +0.05$^\dagger$ & +0.06$^\dagger$ \\
duration & +0.00 & -0.03$^\dagger$ & +0.04$^\dagger$ & +0.04$^\dagger$ \\
words per second & -0.04$^\dagger$ & -0.01 & -0.01 & +0.00 \\
typical voice & -0.00 & -0.01 & +0.05$^\dagger$ & +0.07$^\dagger$ \\
energy, first 50 ms & -0.05$^\dagger$ & -0.03$^\dagger$ & -0.07$^\dagger$ & -0.04$^\dagger$ \\
energy, last 50 ms & -0.15$^\dagger$ & -0.17$^\dagger$ & -0.02 & -0.11$^\dagger$ \\
leading silence & +0.04$^\dagger$ & +0.02 & +0.06$^\dagger$ & +0.02 \\
trailing silence & +0.16$^\dagger$ & +0.16$^\dagger$ & +0.01 & +0.11$^\dagger$ \\
pause ratio & -0.02 & -0.04$^\dagger$ & +0.04$^\dagger$ & +0.03 \\
longest pause & -0.03$^\dagger$ & -0.05$^\dagger$ & +0.05$^\dagger$ & +0.04$^\dagger$ \\
dynamic range & +0.01 & +0.02 & +0.06$^\dagger$ & +0.07$^\dagger$ \\
ends a sentence & -0.01 & -0.00 & -0.03$^\dagger$ & -0.06$^\dagger$ \\
ends with comma & +0.04$^\dagger$ & +0.02 & +0.01 & +0.02 \\
number of words & -0.03$^\dagger$ & -0.03$^\dagger$ & +0.05$^\dagger$ & +0.04$^\dagger$ \\
transcript log-prob, mean & +0.10$^\dagger$ & +0.10$^\dagger$ & +0.03$^\dagger$ & +0.10$^\dagger$ \\
transcript log-prob, min & +0.11$^\dagger$ & +0.11$^\dagger$ & -0.01 & +0.07$^\dagger$ \\
transcript log-prob, end & -0.00 & +0.01 & -0.01 & +0.00 \\
no-speech probability & -0.05$^\dagger$ & -0.04$^\dagger$ & -0.02 & -0.09$^\dagger$ \\
\bottomrule
\end{tabular}
\caption{All 21 prompt descriptors (Table~\ref{tab:descr}): within-recording Spearman correlation with the prompt's intelligibility (fewer word errors = higher), averaged over the 400 test recordings. $^\dagger$: interval excludes zero.}
\label{tab:descriptors}
\end{table*}

\section{Robustness and Ablations}
\label{app:stats}
\label{app:robust}

Table~\ref{tab:extra} collects further results for Qwen3-TTS 0.6B on the 400 test recordings (WER gain over a random prompt in pp unless noted).
Two rows check the prompt transcripts behind the selector's gain.
\emph{Transcript disagreement} is the word error rate between Whisper's and wav2vec~2.0's transcripts of a prompt, a sign of an unreliable transcript; the selector's picks are about as reliable as the UTMOS rule's, both better than an average candidate.
\emph{Consensus prompts} are those whose Whisper transcript and YouTube caption agree; restricting every recording's candidates to them, the selector stays ahead of the rule.
\emph{Caption resynthesis} repeats one test set with the YouTube captions as prompt transcripts.

\begin{table*}[tp]
\centering\small
\begin{tabular}{p{8.4cm}p{6.6cm}}
\toprule
quantity & value \\
\midrule
audio selector $-$ UTMOS rule (pre-specified) & $+0.54$ $[0.14, 0.93]$ \\
same, training seeds 1 and 2 (same recipe and data) & $+0.53$ $[-0.06, 1.03]$; $+0.38$ $[-0.24, 0.90]$ \\
selecting by the mean-WER head instead of the WER rank (not pre-specified) & 1.38 \\
failed outputs excluded: audio selector, UTMOS rule & 0.77, 0.25 \\
wav2vec~2.0 as WER judge: audio selector, UTMOS rule & 1.59, 0.88 \\
prompt-transcript disagreement (Whisper vs.\ wav2vec~2.0) of the pick: audio selector, UTMOS rule, all candidates & 0.23, 0.26, 0.30 \\
share of 3/5/10\,s picks (\%): audio selector, descriptor ranker, UTMOS rule, all candidates & 63/32/5, 67/30/4, 80/18/2, 53/32/16 \\
candidates whose Whisper transcript and YouTube caption agree (24\% of prompts, 307 recordings): audio selector $-$ UTMOS rule, Qwen3-TTS and F5-TTS & $+0.31$, $+0.86$ \\
caption resynthesis (one test set): audio selector $-$ UTMOS rule, Qwen3-TTS and F5-TTS, Whisper vs.\ caption transcripts & $+0.63$ vs.\ $+0.66$; $+1.22$ vs.\ $+1.05$ \\
caption resynthesis: probe selection, Whisper vs.\ caption transcripts & 1.48, 2.35 \\
\bottomrule
\end{tabular}
\caption{Further test results for Qwen3-TTS 0.6B (and F5-TTS where noted).}
\label{tab:extra}
\end{table*}

\begin{table*}[tp]
\centering\small
\begin{tabular}{lrrrr}
\toprule
candidates & audio selector & descriptor ranker & UTMOS rule & audio $-$ rule \\
\midrule
all 19 candidates & 1.36 & 1.00 & 0.83 & +0.54$^\dagger$ \\
only 3\,s prompts & 1.37 & 1.34 & 0.80 & +0.57$^\dagger$ \\
only 5\,s prompts & 0.90 & 0.73 & 0.27 & +0.63$^\dagger$ \\
only 10\,s prompts & 0.64 & 0.46 & 0.06 & +0.58 \\
equal numbers of each length & 1.20 & 0.92 & 0.34 & +0.86$^\dagger$ \\
\bottomrule
\end{tabular}
\caption{WER gain over a random prompt on the test (Qwen3-TTS 0.6B) for different candidate sets: all 19 candidates of a recording (10, 6 and 3 prompts of 3, 5 and 10\,s), candidates of one length only, and equal numbers of each length (30 random draws). $^\dagger$: audio selector better than the UTMOS rule.}
\label{tab:durations}
\end{table*}

\begin{table*}[tp]
\centering\small
\begin{tabular}{lrrr}
\toprule
variant (cross-validation) & WER & change & descriptor ranker \\
\midrule
audio selector (final recipe) & 1.25 &  & 0.90 \\
\quad single model (no ensemble) & 1.20 & -0.05 & 0.93 \\
\quad pointwise loss & 1.18 & -0.07 & 0.90 \\
\quad no comparison with the recording's mean & 1.27 & +0.03 & -- \\
\quad attention across candidates & 1.29 & +0.04 & -- \\
\quad + transcript stream & 1.15 & -0.10 & -- \\
\quad labels: prompt's own UTMOS (no synthesis) & 0.63 & -0.62 & 0.55 \\
\quad labels: prompt's own DNSMOS (no synthesis) & 0.43 & -0.82 & 0.59 \\
\quad 25\% of training recordings & 1.10 & -0.15 & 0.74 \\
\quad 50\% of training recordings & 1.21 & -0.04 & 0.82 \\
\quad labels from one seed (a third of the syntheses) & 1.14 & -0.11 & 0.87 \\
\quad 25\% of recordings, one seed (a twelfth) & 1.11 & -0.14 & 0.78 \\
\quad Whisper-small encoder (88M) & 1.14 & -0.11 & -- \\
\quad Whisper-large-v3 encoder (637M) & 1.22 & -0.03 & -- \\
\quad WavLM-base+ encoder (94M) & 1.24 & -0.00 & -- \\
\quad CLAP audio encoder (69M) & 0.42 & -0.82 & -- \\
\bottomrule
\end{tabular}
\caption{All ablations by cross-validation on the training recordings (folds keep same-voice recordings together; mean of three seeds): WER gain over a random prompt, its change from the final recipe (differences below about 0.1 are within the spread between seeds), and the descriptor ranker trained with the same labels on the same folds (--: the variant does not concern it).}
\label{tab:ablations_all}
\end{table*}

\paragraph{Prosody.}
For Qwen3-TTS 0.6B on the test, we tracked F0 (Praat, 10\,ms steps) in every output and in the speaker's own reading of each target sentence, and measured an output's pitch variability (standard deviation of F0 in semitones), its absolute difference from that of the speaker's reading, and its tempo error (absolute log ratio of the voiced durations of output and reading).
With a random prompt these are 3.67\,st, 1.78\,st and 0.29.
All three selections lower pitch variability slightly (audio selector $-0.06$ $[-0.11, -0.01]$\,st, UTMOS rule $-0.07$ $[-0.12, -0.02]$, DNSMOS rule $-0.04$ $[-0.09, +0.01]$), the audio selector no more than the UTMOS rule (difference $+0.01$ $[-0.06, +0.08]$).
Measured against the speaker's reading, the audio selector matches its pitch variability as well as a random prompt does (error $-0.01$ $[-0.05, +0.03]$\,st) and its tempo slightly better (error $-0.013$ $[-0.019, -0.006]$).
These measures cover global prosody only, not emphasis or context-appropriate stress.

\section{Quality Rules}
\label{app:rules}

Table~\ref{tab:rules_app} gives the two quality rules on the 400 training recordings of \S\ref{sec:sensitivity} for all three TTS models, among the 3\,s candidates as in Table~\ref{tab:sensitivity}, which gives probe selection's gains.

\begin{table*}[tp]
\centering\small
\begin{tabular}{lrrrrrr}
\toprule
rule & WER & failures & UTMOS & DNSMOS & Audiobox & similarity \\
\midrule
\multicolumn{7}{l}{\textit{Qwen3-TTS 0.6B}} \\
UTMOS rule & 0.37 & 0.33 & 0.078 & 0.022 & 0.046 & 0.029 \\
DNSMOS rule & 0.18 & 0.43 & 0.065 & 0.033 & 0.034 & 0.023 \\
\midrule
\multicolumn{7}{l}{\textit{Chatterbox}} \\
UTMOS rule & 0.12 & -0.10 & 0.085 & 0.028 & 0.053 & 0.024 \\
DNSMOS rule & 0.18 & 0.01 & 0.071 & 0.040 & 0.056 & 0.020 \\
\midrule
\multicolumn{7}{l}{\textit{F5-TTS}} \\
UTMOS rule & 1.31 & 1.27 & 0.083 & 0.021 & -0.005 & 0.032 \\
DNSMOS rule & 1.79 & 1.95 & 0.056 & 0.026 & 0.014 & 0.029 \\
\bottomrule
\end{tabular}
\caption{Quality rules on 400 training recordings, among the 3\,s candidates: gain over a random prompt (probe selection in Table~\ref{tab:sensitivity}).}
\label{tab:rules_app}
\end{table*}

\begin{table*}[tp]
\centering\small
\begin{tabular}{lrrrrr}
\toprule
TTS model & UTMOS rule & DNSMOS rule & best descriptor & audio selector & probe selection \\
\midrule
Qwen 0.6B & 0.83 & 0.33 & 0.96 & 1.36$^{a,b}$ & 1.35 \\
Qwen 1.7B$^*$ & 0.28 & -0.25 & 0.44 & 1.49$^{a,b,c}$ & 1.86 \\
Chatterbox & 0.03 & -0.01 & 0.13 & 0.25$^{b}$ & 0.30 \\
F5-TTS & 0.97 & 1.19 & 0.66 & 2.64$^{a,b,c}$ & 3.00 \\
\bottomrule
\end{tabular}
\caption{WER gain over a random prompt on the 400 test recordings (Figure~\ref{fig:models}). The best descriptor is chosen per model on its training recordings ($^*$: Qwen3-TTS 1.7B has no training outcomes and uses the 0.6B choice and selector). Superscripts on the selector: better than the UTMOS rule (a), the DNSMOS rule (b) or the best descriptor (c).}
\label{tab:baselines}
\end{table*}

\begin{table*}[tp]
\centering\small
\begin{tabular}{lrrrrrr}
\toprule
& \multicolumn{2}{c}{UTMOS} & \multicolumn{2}{c}{Audiobox} & \multicolumn{2}{c}{similarity} \\
\cmidrule(lr){2-3}\cmidrule(lr){4-5}\cmidrule(lr){6-7}
TTS model, selector & selector & rule & selector & rule & selector & rule \\
\midrule
Qwen3-TTS 1.7B, selector trained on 0.6B & 0.054$^\ddagger$ & 0.092 & 0.011 & 0.038 & -0.000 & -0.005 \\
Chatterbox, own selector & 0.046 & 0.069 & 0.029 & 0.048 & 0.005$^\dagger$ & -0.008 \\
Chatterbox, selector trained on Qwen 0.6B & 0.039$^\ddagger$ & 0.069 & -0.003$^\ddagger$ & 0.048 & -0.005 & -0.008 \\
F5-TTS, own selector & 0.052$^\ddagger$ & 0.103 & 0.030 & 0.045 & 0.000 & -0.009 \\
F5-TTS, selector trained on Qwen 0.6B & 0.021$^\ddagger$ & 0.103 & 0.006 & 0.045 & -0.001 & -0.009 \\
\bottomrule
\end{tabular}
\caption{Other TTS models on the 400 test recordings: gain over a random prompt of the audio selector and the UTMOS rule on naturalness and similarity (WER in Table~\ref{tab:baselines} and \S\ref{app:models}). $^\dagger$/$^\ddagger$: selector better/worse than the rule.}
\label{tab:other}
\end{table*}

\section{Other TTS Models}
\label{app:models}

\paragraph{Agreement between models.}
For a metric, a prompt's outcome under model $a$ is its mean over its 15 outputs.
The agreement $\rho_{ab}$ of two models is the Spearman correlation between their outcomes over the candidates of a recording, averaged over the 400 test recordings.
It is limited by how reproducible each model's outcomes are.
We therefore split every prompt's outputs into two halves (alternating combinations of target sentence and seed, 7--8 outputs each), compute the same within-recording Spearman correlation between the means of the two halves, average it over recordings ($r^{\mathrm{half}}_a$), and take the Spearman--Brown reliability of the full mean, $r_a = 2r^{\mathrm{half}}_a/(1+r^{\mathrm{half}}_a)$.
The corrected agreement (Figure~\ref{fig:agreement}, right) is
\[ \rho^{\mathrm{corr}}_{ab} = \frac{\rho_{ab}}{\sqrt{r_a\, r_b}}, \]
computed after averaging over recordings.
Averaging Fisher-$z$-transformed correlations instead changes the corrected values by at most 0.04 and leaves their order unchanged.
The reliabilities range from 0.29 (Chatterbox, WER) to 0.97 (F5-TTS, similarity), none near zero; for WER they are 0.52 and 0.59 (Qwen3-TTS 0.6B and 1.7B), 0.29 (Chatterbox) and 0.64 (F5-TTS).
Uncorrected, the agreement on WER is 0.35 between the two Qwen models, 0.22--0.23 between them and F5-TTS, and at most 0.03 with Chatterbox.

\paragraph{Transfer.}
Figure~\ref{fig:agreement} (left) gives transfer in percent of probe selection's gain; in WER gain (pp), the selector trained on Qwen3-TTS 0.6B gains 1.49, 0.14 and 1.53 on Qwen3-TTS 1.7B, Chatterbox and F5-TTS, the one trained on Chatterbox 0.88, 0.77 and 1.77 on Qwen3-TTS 0.6B, 1.7B and F5-TTS, and the one trained on F5-TTS 1.04, 1.57 and 0.23 on Qwen3-TTS 0.6B, 1.7B and Chatterbox.
Table~\ref{tab:baselines} gives each model's WER gains and Table~\ref{tab:other} the other metrics.

\section{Prompts Differing in Noise}
\label{app:noise}

From the first 20 speakers of LibriSpeech dev-clean we cut 8 candidate prompts each of 3, 5 and 10\,s and add white noise or DEMAND noise \citep{thiemann2013demand} at 20, 10 and 5\,dB SNR; each selection draws every candidate's condition at random among the clean version and these six, and all candidates of a speaker compete in one selection (200 draws per speaker).
Prompt transcripts are the LibriSpeech references; outputs come from Qwen3-TTS 1.7B with the same five target sentences and three seeds per prompt as on YouTube.
Gains are averaged over the draws of a speaker, and intervals resample the 20 speakers, not the draws.

\section{Listening Test}
\label{app:human}

\paragraph{Design.}
Qwen3-TTS 0.6B on the two test sets; among the recordings where the audio selector and the UTMOS rule pick different prompts (excluding recordings used in a pilot), 72 were drawn at random, each with one random target sentence and sampling seed, the same for both outputs.
Each listener rated half of the pairs (about 15 minutes), after one practice pair, with three attention checks mixed in: the same sentence and seed from a failed output (WER of at least 80\%) and a perfect one.
Order and sides were randomized per listener.
For each pair, listeners answered ``Which says the sentence more accurately and clearly?'' and ``Which sounds more natural?'' on a five-point scale (much better, slightly better, about the same, for either side), scored from $+2$ to $-2$ in favour of the selector.
The protocol was fixed before any rating, including the exclusion rule (a listener who does not prefer the correct output on clarity in at least two of the three checks); no listener was excluded.

\paragraph{Results.}
Each pair has 2--5 ratings; we average a pair's ratings and then the pairs, so that every pair counts equally; intervals resample both pairs and listeners.
Clarity: mean $+0.02$ $[-0.20, 0.23]$; naturalness: mean $+0.08$ $[-0.16, 0.31]$.
By the automatic WER of the two outputs, the selector's output has fewer errors in 12 pairs, the rule's in 11, and both are free of errors in 44.
Mean clarity is $+0.39$ where the selector's output has fewer errors and $-0.14$ where the rule's has.
Agreement between listeners (Krippendorff's $\alpha$) is 0.33--0.69 for clarity and 0.17--0.23 for naturalness.

\paragraph{What one output per pair can show.}
Over all test recordings where the two prompts differ, and all of their sentences and seeds, Whisper finds the selector's output better in 20\% of the output pairs, the rule's in 13\%, and both equal in 67\%.
A listener who agreed with Whisper on every pair ($+1$, $0$ or $-1$) would thus score $+0.06$ on average with a standard deviation of about $0.6$ per pair, so its 95\% interval excludes zero only from about 305 pairs; real listeners, who tie more often and disagree with each other, need more.
In the 23 test pairs where the outputs differ in word errors, 62\% of the ratings that prefer a side prefer the side with fewer errors by Whisper (mean clarity in Whisper's direction $+0.25$, interval $[-0.29, 0.81]$).

\end{document}